\documentclass[prb,aps,twocolumn]{revtex4}
\usepackage{graphicx,amssymb,amsmath,color,psfrag}
\usepackage{amsthm}
\usepackage{amsfonts}
\usepackage{algorithmic}
\usepackage{enumerate}
\usepackage{latexsym}

\begin{document}
\newcommand{\dt}{\Delta\tau}
\newcommand{\al}{\alpha}
\newcommand{\ep}{\varepsilon}
\newcommand{\ave}[1]{\langle #1\rangle}
\newcommand{\have}[1]{\langle #1\rangle_{\{s\}}}
\newcommand{\bave}[1]{\big\langle #1\big\rangle}
\newcommand{\Bave}[1]{\Big\langle #1\Big\rangle}
\newcommand{\dave}[1]{\langle\langle #1\rangle\rangle}
\newcommand{\bigdave}[1]{\big\langle\big\langle #1\big\rangle\big\rangle}
\newcommand{\Bigdave}[1]{\Big\langle\Big\langle #1\Big\rangle\Big\rangle}
\newcommand{\braket}[2]{\langle #1|#2\rangle}
\newcommand{\up}{\uparrow}
\newcommand{\dn}{\downarrow}
\newcommand{\bb}{\mathsf{B}}
\newcommand{\ctr}{{\text{\Large${\mathcal T}r$}}}
\newcommand{\sctr}{{\mathcal{T}}\!r \,}
\newcommand{\btr}{\underset{\{s\}}{\text{\Large\rm Tr}}}
\newcommand{\lvec}[1]{\mathbf{#1}}
\newcommand{\gt}{\tilde{g}}
\newcommand{\ggt}{\tilde{G}}
\newcommand{\jpsj}{J.\ Phys.\ Soc.\ Japan\ }

\title{Magnetic Impurity Affected by Spin-Orbit Coupling: Behavior near a Topological Phase Transition}
\author{F. M. Hu$,^{1}$ T. O. Wehling$,^{2,3}$ J. E. Gubernatis$,^{4}$ Thomas Frauenheim$,^{3}$ R. M. Nieminen$^1$}
\affiliation{$^1$COMP/Department of Applied Physics, Aalto University School of Science, P.O. Box 11100, FI-00076 Aalto, Espoo, Finland\\
$^{2}$ Institute of Theoretical Physics, University of Bremen, Otto-Hahn-Allee 1, D-28359 Bremen, Germany\\
$^{3}$ Bremen Center for Computational Materials Science, University of Bremen, Am Fallturm 1a, D-28359 Bremen, Germany\\
$^{4}$ Theoretical Division, Los Alamos National Laboratory, Los Alamos, New Mexico 87545, USA}

\begin{abstract}
We investigate the effect of spin-orbit coupling on the behavior of magnetic impurity at the edge of a zigzag graphene ribbon by means of quantum Monte Carlo simulations. A peculiar interplay of Kane-Mele type spin-orbit and impurity-host coupling is found to affect local properties such as the impurity magnetic moment and spectral densities. The special helical nature of the topological insulator on the edge is found to affect nonlocal quantities, such as the two-particle and spin-spin correlation functions linking electrons on the impurity with those in the conduction band.
\end{abstract}

\date{\today}
\maketitle

\section{Introduction}
Spin-orbit coupling (SOC) plays a central role in topological insulator (TI) materials, \cite{RMP11} it opens a gap inside the bulk but supports the gapless electron states on the boundary. On the boundary of TI, impurity potential scattering is restricted by time-reversal symmetry: \cite{Wu06,Xu06} the backscattering among the electrons is allowed only if it is accompanied by a spin flip. This restriction gives rise to novel Kondo physics of magnetic impurities in TI. \cite{Maciejko09,Tanaka11} Accordingly, the behavior of magnetic impurities in host materials with SOC has recently attracted interest. In two-dimensional (2D) systems, several theoretical papers \cite{Feng10,Isaev12} report that the presence of the SOC can in general protect impurity's magnetic moment from being totally screened. However, for the specific case of the Rashba coupling \cite{Zitko11,Zarea12} it has been reported that the SOC only makes a small or a high-order contribution to Kondo scaling.

In general, SOC and energies associated with the Kondo screening of magnetic impurities can be of the same order of magnitude. It remains to be explored what effects arise when crossing over from weak to strong SOC. In particular, the question of how the impurity behaves when the host undergoes a topological phase transition from a normal state to TI is an open question.

In this paper, we consider an Anderson impurity on the edge of a zigzag graphene ribbon with a Kane-Mele type SOC gradually strengthened, so that a topological phase transition occurs in the system. Using quantum Monte Carlo (QMC) simulations at finite temperature, we perform a comprehensive study on the properties of impurity. Calculating local physical quantities on the impurity site, such as the average double occupancy, magnetic moment, spin susceptibility and spectral densities, we find that in general the SOC can support the local moment formation. We see that in the case of zero or weak SOC,  impurity is mainly dominated by its coupling with localized edge states, realizing a situation similar to the zero band width Anderson (ZBWA) model. In the case of strong SOC, the edge states are greatly broadened and the local density of states (LDOS) is suppressed,  so that the impurity behaves as a spin in a normal metal. This difference between strong and weak coupling also manifests itself in the dependence of the local moment on chemical potential, which in graphene can be tuned by a gate voltage. We also study nonlocal linking of the electrons on impurity with those in the conduction band. In nonlocal two-particle correlations, we observe a set of distinct spin-momentum relations which show the helical locking and interplay between backscattering and spin flip. This interplay  signals the appearance of a topological phase in the host. Additionally, we find the spin rotation symmetry in Kondo cloud around impurity to be partially broken by the SOC. This finding agrees with a previous study of an Anderson impurity in 2D helical metal with variational method that is valid for large Coulomb interaction. \cite{Feng10} In this work, we study the effect of electron-electron interaction over a wide regime and document that the Coulomb interaction enhances the anisotropy in spin-spin correlation function and plays a complicated role in spin and momenta scattering.

\section{Model and Methods}
Our starting point is the Hamiltonian
\begin{equation}\label{Eq:KMA}
H=H_{\textrm{K-M}}+H_{1}+H_{2},
\end{equation}
where
$H_{\textrm{K-M}}$ is the Kane-Mele model for a zigzag edge graphene ribbon. \cite{Kane05} It has two pieces:
$H_{\textrm{K-M}}=H_{t}+H_{\textrm{so}}$,
with $H_{t}$ being the usual nearest-neighbor hopping of tight-binding model in graphene
\[
H_{t}=-t\sum_{<ij>,\sigma}(c^{\dag}_{i,\sigma}c_{j,\sigma}+\textrm{H.c.})-\mu\sum_{i,\sigma}c^{\dag}_{i,\sigma}c_{i,\sigma},
\]
and
\[
H_{\textrm{so}}=\lambda\sum_{<<ij>>}(i\nu_{ij}c^{\dag}_i\sigma^zc_{j}+\textrm{H.c.})
\]
$H_{\textrm{so}}$ is the SOC term and $\sigma^{z}$ is the $z$ Pauli matrix. $H_{\textrm{so}}$ thus has opposite signs for opposite electron spins.
The parameters $\nu_{ij}=-\nu_{ji}=\pm1$ depend on the orientation of the two nearest neighbor bonds as the electron hops from site $i$ to $j$: $\nu_{ij}=+1$ if the
electron makes a left turn to the second bond. It is negative if it makes a right turn.
$H_1$ is the impurity Hamiltonian
\[
H_{1}=\sum_{\sigma}(\varepsilon_{d}-\mu)d^{\dag}_{\sigma}d_{\sigma}^{}+Ud^{\dag}_{\uparrow}d_{\uparrow}^{}d^{\dag}_{\downarrow}d_{\downarrow}^{}\texttt{.}
\]
Here $\varepsilon_{d}$ is the energy of the impurity orbital and $U$ is Coulomb repulsion  inhibiting the simultaneous occupancy of the orbital by two electrons. Finally $H_{2}$ describes the hybridization between impurity and one of atoms on the edge (located at A-sublattice site $R_{a0}$),
\[
H_2=V\sum_{\sigma}[c^{\dag}_{a0,\sigma}d_{\sigma}^{}+d^{\dag}_{\sigma}c_{a0,\sigma}^{}]\texttt{.}
\]

Our principal computational tools are the single-impurity QMC algorithm \cite{Hirsch86} for computing the local thermodynamic properties of the impurity and the method of Bayesian statistical inference for computing spectral densities.
The QMC naturally returns the imaginary-time Green's function $G_{d\sigma}(\tau > 0)= \langle d_{\sigma}(\tau)d_{\sigma}^{\dag}\rangle $ of the impurity. With this Green's function, we can easily compute the magnetic quantities on the impurity site
such as the expected values of magnet moment square $\langle(S^{z})^2\rangle=\langle(d_{\uparrow}^{\dag}d_{\uparrow}-d_{\downarrow}^{\dag}d_{\downarrow})^2\rangle$, the double occupancy $\langle n_{d\uparrow}n_{d\downarrow}\rangle=\langle d_{\uparrow}^{\dag}d_{\uparrow}d_{\downarrow}^{\dag}d_{\downarrow}\rangle$
and the static impurity spin susceptibility
\begin{equation}\label{eq:susceptibility}
\chi=\int^{\beta}_{0}d\tau\langle S^{z}(\tau)S^{z}(0)\rangle,
\end{equation}
where $\beta=T^{-1}$, $S^{z}(\tau)=e^{\tau H}S^{z}(0)e^{-\tau H}$. Computing the imaginary-time Green's function also enables us to compute the spectral density $A(\omega)=\sum_\sigma A_\sigma(\omega)$ by numerically solving \cite{jarrell96}
\[
G_d\left( \tau  \right)  ={\int\limits_{ - \infty }^\infty  {d\omega } } \frac{e^{  -\tau\omega}{A  \left(\omega  \right)}}
{{e^{ - \beta \omega}  + 1}}.
\]
Using an extended QMC algorithm, \cite{Gubernatis87} we can calculate the Green's function linking electrons on the impurity site and those in the conduction bands: $G_{di\sigma}(\tau > 0)= \langle d_{\sigma}(\tau)c_{i\sigma}^{\dag}\rangle $ or $G_{id\sigma}(\tau > 0)= \langle c_{i\sigma}(\tau)d_{\sigma}^{\dag}\rangle $.

\section{In the large-$U$ limit: s-d exchange model}

Before we do the QMC simulation, to gain insight of this problem, we map the original Hamiltonian, Eq.~(\ref{Eq:KMA}), to impurity's single-occupancy subspace in the large -$U$ limit. Defining projection operators, \cite{Hewson}
\begin{eqnarray}
P_{0}&=&(1-n_{d\uparrow})(1-n_{d\downarrow}),\nonumber\\
P_{1}&=&n_{d\uparrow}(1-n_{d\downarrow})+n_{d\downarrow}(1-n_{d\uparrow}),\nonumber\\
P_{2}&=&n_{d\uparrow}n_{d\downarrow}.
\end{eqnarray}
and we can solve the effective Hamiltonian in single-occupancy subspace as
\begin{equation}
\tilde{H}=H_{11}+H_{12}(E-H_{22})^{-1}H_{21}+H_{10}(E-H_{00})^{-1}H_{01},
\end{equation}
here $H_{ij}=P_{i}HP_{j}$ and $H_{ij}=H_{ji}^{\dag}$, then an effective s-d exchange model has the formula as
\begin{eqnarray}\label{Eq:sdmodel}
\tilde{H}&=&\frac{1}{N}\sum_{kk'll'}[S^{z}(J^{\perp}_{kk'll'\uparrow}c^{\dag}_{kl\uparrow}c_{k'l'\uparrow}-J^{\perp}_{kk'll'\downarrow}c^{\dag}_{kl\downarrow}c_{k'l'\downarrow})]\nonumber\\
&+&[J^{\parallel}_{kk'll'}S^{+}c_{kl\downarrow}^{\dag}c_{k'l'\uparrow}+\textrm{H.c.}],\nonumber\\
J^{\perp}_{kk'll'\sigma}&=&V_{kl\sigma}^{a}V_{k'l'\sigma}^{a*}\left[\frac{V^2}{U+\varepsilon_d-E_{k'l'}}+\frac{V^2}{E_{kl}-\varepsilon_d}\right],\nonumber\\
J^{\parallel}_{kk'll'}&=&V_{kl\downarrow}^{a}V_{k'l'\uparrow}^{a*}\left[\frac{V^2}{U+\varepsilon_d-E_{k'l'}}+\frac{V^2}{E_{kl}-\varepsilon_d}\right],
\end{eqnarray}
in which $S^{+}=d^{\dag}_{\uparrow}d_{\downarrow}$, and $V_{kl\sigma}^{a}$ is the amplitude of eigenstate at $R_{a0}$ on the edge, here $k$ and $l$ are indexes of momentum and band, respectively. Furthermore, $V_{kl\sigma}^{a}$ relates the Fermi operators in real space to those in eigen space,
\begin{equation}\label{Eq:lineartr}
c_{ar,\sigma}^{\dag}=\frac{1}{\sqrt{N}}\sum_{kl}e^{-ikr}V_{kl\sigma}^{a}c^{\dag}_{kl\sigma},
\end{equation}
$N$ is the number of atoms on the edge. Therefore we see that $\tilde{H}$ is different from the normal s-d exchange model because of the broken symmetry,
$J^{\perp}_{kk'll'\sigma}\neq J^{\parallel}_{kk'll'}$, which is due to the helical properties driven by the SOC. Later we will see this asymmetry more clearly in the spin-flip scattering. If we consider that the impurity is mainly scattered by the levels near the Fermi level and $-\varepsilon_{d}=\frac{U}{2}\gg|\mu|$, there are $J^{\perp}_{kk'll'\sigma}\approx\frac{4V^2}{U}V_{kl\sigma}^{a}V_{k'l'\sigma}^{a*}$ and $J^{\parallel}_{kk'll'}\approx\frac{4V^2}{U}V_{kl\downarrow}^{a}V_{k'l'\uparrow}^{a*}$

\begin{figure}[t]
\begin{center}
\includegraphics[scale=0.4,bb=5 254 665 502]{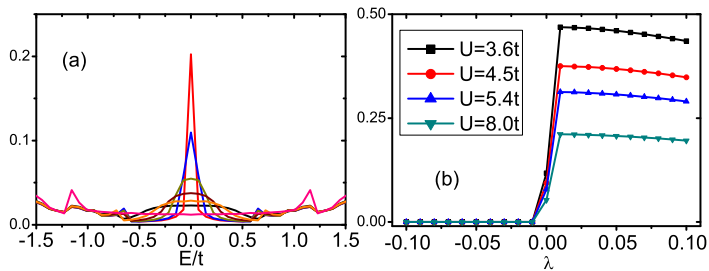}
\end{center}
\caption{(color online). (a) The LDOS as a function of energy $E/t$ with different values for $\lambda$. From top to bottom $\lambda =$ 0, $0.02t$, $0.04t$, $0.06t$, $0.08t$, $0.1t$ and $0.2t$.
(b) The spin-flip strength $J^{\parallel}_{k-kl}$ as a function of $\lambda$.}\label{Fig:sd}
\end{figure}

In Hamiltonian~(\ref{Eq:sdmodel}), the total amplitude of spin exchange depends on the sum $\sum_{kk'll'}$, in particular, the elastic exchange is determined by the sum  $\sum_{kk'l}$, which is just the density of states (the degeneracy) on the level $l$. Fig.~\ref{Fig:sd}(a) shows the LDOS on the edge and we see that when $\lambda=0$, a sharp peak exists at the Dirac point ($E/t=0$) suggesting the existence of a strongly localized edge states. When $\lambda \neq 0$, the central peaks becomes smooth, and at $\lambda=0.2t$, their values near $E/t=0$ approach a constant and thus near the Dirac point become similar to the LDOS in a normal metal. First, we can expert that since the SOC decreases the LDOS on the edge, the spin exchange between impurity and conduction electrons will be suppressed.

Second, we focus our attention on a special process, elastic backscattering accompanied by spin flip between Kramers pair ($\pi\pm k, \sigma\bar{\sigma}$), in the following we denote this pair as ($k\sigma, -k\bar{\sigma}$). In this process, the transversal exchange $J^{\parallel}_{k-kl}$ is nonzero, but due to the helical phase on the edge, it has no corresponding vertical term in $\tilde{H}$, and $J^{\perp}_{k-kl\sigma}=0$. Thus the symmetry in the s-d exchange model (\ref{Eq:sdmodel}) is broken by the SOC.  The weight of this spin-flip scattering, $S^{+}c_{kl\downarrow}^{\dag}c_{-kl\uparrow}$, depends on the coupling strength $J^{\parallel}_{k-kl}$, which is shown as the function of $\lambda$ in Fig.~\ref{Fig:sd}(b). Here the momentum $k=0.0025\pi$ and $l$ is the first band under zero point. We see that $J^{\parallel}_{k-kl}$ is totally equal to zero in negative regime of $\lambda$ but in positive regime has finite value. This is due to helical properties on the edge: with positive SOC, the spin-up and spin-down electrons have momenta $-k$ and $k$, respectively, so both $V^{a}_{k\downarrow}$ and $V^{a*}_{-k\uparrow}$ have non-zero values but vanishes when $\lambda$ changes the sign and the electrons will move along oppositive directions. Moreover, in Fig.~\ref{Fig:sd}(b) we see that the strength of spin flip $J^{\parallel}_{k-kl}$ decreases as the SOC increases in the large-$U$ limit.


\section{Numerical results}
\subsection{Local properties of impurity}
\begin{figure}
\begin{center}
\includegraphics[scale=0.40, bb=128 55 437 458]{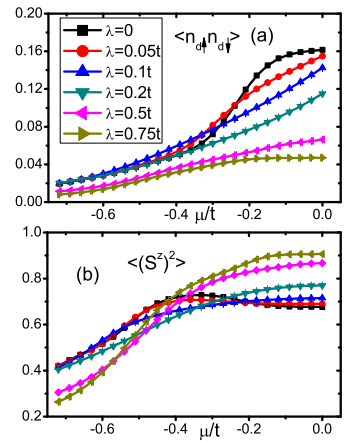}
\end{center}
\caption{(color online). (a) Double occupancy $\left<n_{d\uparrow}n_{d\downarrow}\right>$ versus chemical potential $\mu$. (b) $\left<(S^{z})^2\right>$ versus $\mu/t$.  Here, $U=1.2t$, $V=0.65t$, $\varepsilon_{d}=-U/2$, and $T^{-1}=32t^{-1}$. }\label{Fig:tdq}
\end{figure}

In this section, we will display our QMC results for both local and nonlocal properties for impurity. To see the behavior of local magnetic moment, in Fig.~\ref{Fig:tdq}, we show double occupancy $\left<n_{d\uparrow}n_{d\downarrow}\right>$ and local moment square $\left<(S^{z})^2\right>$ versus the chemical potential $\mu$ which can be tuned by an electric field in graphene.
With $\varepsilon_{d}=-U/2$ fixed and $\mu$ is near zero point the impurity site is half filled, and thus
the magnetic moment is driven by avoiding the possibility of double occupancy. In Fig.~\ref{Fig:tdq}(b) we see that increasing the SOC results in the average double occupancy $\left<n_{d\uparrow}n_{d\downarrow}\right>$ decreasing because the SOC suppresses the effective hybridization between
impurity and edge states. Near the Dirac point, the average local magnetic moments  $\left<(S^{z})^2\right>$ is enhanced by the SOC. We also note that with weak SOC, i.e.,
$0.0< \lambda< 0.05t$, the maxima of $\left<(S^{z})^2\right>$ are not at $\mu=0$ while for the large $\lambda$,
they are at $\mu=0$. We comment that with small SOC the localized states on the edge are antiferromagnetically coupled
to impurity states below the Fermi energy, so shifting $\mu$ from zero can decouple these oppositely-aligned spin states and lead to the development of a magnet moment.  \cite{hu2012}
Contrary to the weak SOC case, the strong SOC greatly broadens the central peak in LDOS on the edge, making the LDOS near the Fermi energy similar to that of a normal metal. Thus we see that half filling (hole-particle symmetry) optimizes the magnetic moment at $\mu=0$.

\begin{figure}
\begin{center}
\includegraphics[scale=0.33, bb=90 44 562 410]{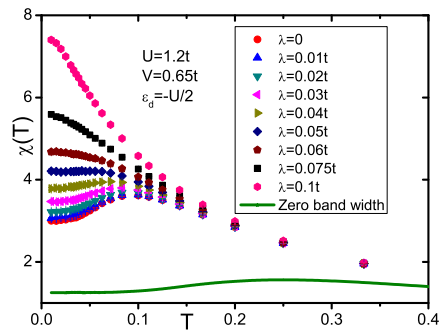}
\end{center}
\caption{(color online). The spin susceptibility $\chi$ as a function of $T$ for different values of the SOC.
$U=1.2t$, $V=0.65t$ $\varepsilon_d=-U/2$ and $\mu=0$. The solid line presents the results for ZBWA model.}\label{Fig:Tchi}
\end{figure}

In order to see the formation of local moment and screening around it, we study the dynamical behavior of spin susceptibility in Eq.~(\ref{eq:susceptibility}), in figure~\ref{Fig:Tchi} we shows $\chi$ as a function of temperature for different values of $\lambda$. We fix the  hole-particle symmetry by setting $\mu=0$ and $\varepsilon_d=-U/2$. Doing so means the average electron
occupancy of the impurity site is one; i.e., it is half filled.
From the figure, we also see that with small $0<\lambda<0.05t$ and the lowing of temperature, the spin susceptibility first increases, then decreases and finally saturates (totally screened).
But when $0.05t<\lambda<0.1t$, $\chi$ first increases and then directly becomes saturated. We note that when the SOC is gradually switched on, $\chi$ goes cross over form the behavior in ZBWA model (solid line in figure~\ref{Fig:Tchi}) to that in a normal metal. \cite{Hewson} We propose this transition occurs because of the LDOS decreased by the SOC, consequently, the spin exchange between impurity and conduction electrons is suppressed.

\begin{figure}[t]
\begin{center}
\includegraphics[scale=0.30, bb=37 24 640 586]{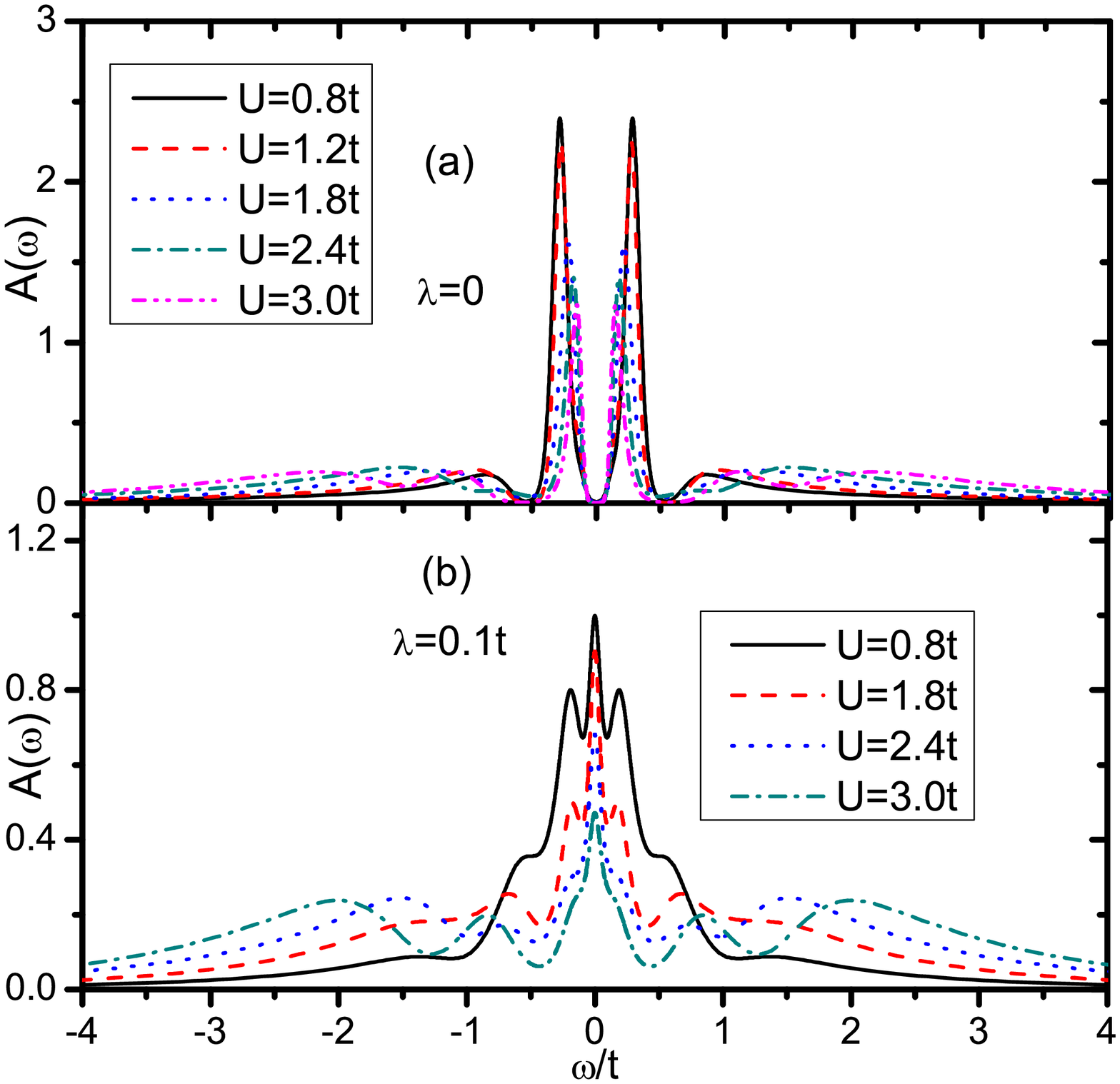}
\end{center}
\caption{(color online). The spectral density $A(\omega)$ versus $\omega/t$. (a) The SOC $\lambda=0$. (b) The SOC $\lambda=0.1t$. In both figures,
$V=0.65t$, $\mu=0$, $\varepsilon_{d}=-U/2$ and $T^{-1}=32t^{-1}$. }\label{Fig:Aw}
\end{figure}

In Fig.~\ref{Fig:Aw}, we show additional spectral densities $A(\omega)$ for the impurity site. Here we have
$A(\omega)=A(-\omega)$ due to hole-particle symmetry. In Fig.~\ref{Fig:Aw}a are results for $\lambda=0$.
In this case the impurity states are strongly correlated with the edge states. There are four peaks.
The inner two stem from the localized edge states which are split by the impurity state. The outer two peaks stem from the impurity level which are split by the Coulomb interaction into $\varepsilon_d$ and $\varepsilon_d+U$.
This four-peak structure is similar to that in ZBWA model. It is known that in ZBWA model the separations and weights of two inner peaks are proportional to $U^{-1}$ and $U^{-2}$, respectively, \cite{Hewson}
 but here we cannot see this dependence on $U$.

In Fig.~\ref{Fig:Aw}b, we show $A(\omega)$ for $\lambda=0.1t$. Clearly visible is the central Kondo resonance and the heights of two inner peaks decreasing with increasing $\lambda$. This behavior is consistent with the expectation that increasing $\lambda$ broadens the central peak in LDOS, making a singularity become flat band.  We also see that when we increase $U$, the height of the central Kondo peak, as well as that of the two inner peaks, decrease, but the heights of the outer two peaks from impurity levels increase. These behaviors are similar to those of an impurity in a normal metal. Additionally, the two smooth peaks around $\omega=\pm t$ are remnants of van Hove singularities.

From figures~\ref{Fig:tdq} to~\ref{Fig:Aw}, we mainly show the local properties on impurity site, which are exactly governed by the edge states modified by the SOC. The central resonant in LDOS is broadened by the SOC, and the region near the Dirac point approach a constant. So as the SOC is increasing in the system, the impurity coupled to the edge states crosses over from a spin in a band with zero width to one in a normal metal. However, these local physical quantities cannot reflect the natures of quantum spin Hall states, and in the following part, we will display the non-local correlation between impurity and conduction electrons, which directly characterizes helical phase on the edge.

\subsection{Nonlocal properties of impurity}
\begin{figure}
\begin{center}
\includegraphics[scale=0.43, bb=53 540 545 766]{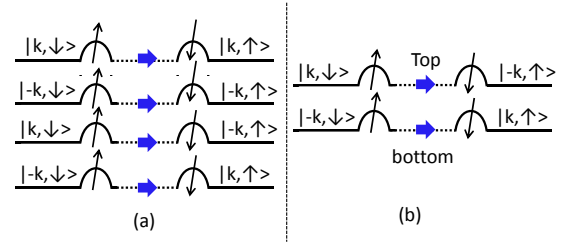}
\end{center}
\caption{(color online). (a) The spin-flip processes in the normal states. (b) The spin-flip processes on the helical edges. $|\pm k,\uparrow\downarrow>$ are the states of conduction electrons and up and down arrows mean local spins.}\label{Fig:schematic}
\end{figure}

\begin{figure}
\begin{center}
\includegraphics[scale=0.38, bb=8 155 629 408]{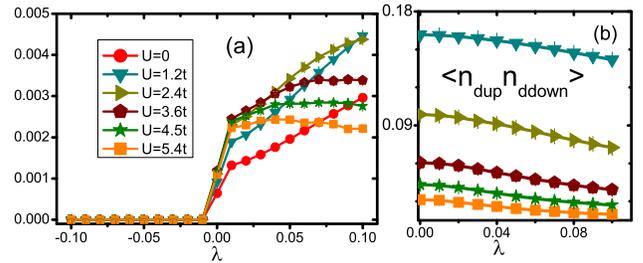}
\end{center}
\caption{(color online). (a) Correlation function for backscattering and spin flip, $\langle c_{-k,\uparrow}d_{\uparrow}^{\dag}d_{\downarrow}c_{k,\downarrow}^{\dag}\rangle$ versus $\lambda$, (b) double occupancy as a function of $\lambda$.
In both figures $V=0.65t$, $\varepsilon_d=-U/2$, $\mu=0$, $\beta=32t^{-1}$ and momentum $k=0.00125\pi$.}\label{Fig:kcor1}
\end{figure}


The most interesting property of a TI boundary is time-reversal invariance causing backscattering always being accompanied by spin-flip scattering.
In Fig.~\ref{Fig:schematic}, the spin-flip processes in the normal states and helical phase are shown. In the normal states, both forward and back spin-flip scatterings are allowed on the edge, while in the helical phase, on each edge, only one process exits for right or left mover with fixed spin orientation.
In order to directly see the signatures of spin-flip scattering, we computed a set $P_k$ of two-particle correlation functions between the impurity and Kramers pair with momentum $k$: $P_k=\{
\langle c_{ k,\uparrow}d_{\uparrow}^{\dag}d_{\downarrow}c_{-k,\downarrow}^{\dag}\rangle,
\langle c_{-k,\uparrow}d_{\uparrow}^{\dag}d_{\downarrow}c_{ k,\downarrow}^{\dag}\rangle,
\langle c_{ k,\uparrow}d_{\uparrow}^{\dag}d_{\downarrow}c_{ k,\downarrow}^{\dag}\rangle,$
$\langle c_{-k,\uparrow}d_{\uparrow}^{\dag}d_{\downarrow}c_{-k,\downarrow}^{\dag}\rangle
\}$.  When $\lambda\neq0$, due to the helical properties on the edge, with positive $\lambda$ the correlation function $\langle c_{-k,\uparrow}d_{\uparrow}^{\dag}d_{\downarrow}c_{k,\downarrow}^{\dag}\rangle$ is nonzero but with negative $\lambda$ it is always equal to zero. For $\langle c_{k,\uparrow}d_{\uparrow}^{\dag}d_{\downarrow}c_{-k,\downarrow}^{\dag}\rangle$, the situation is completely opposite. As for forward scattering with spin flip, when the system in helical liquid phase, these processes are forbidden, namely, when $\lambda\neq0$, the correlation functions $\langle c_{ k,\uparrow}d_{\uparrow}^{\dag}d_{\downarrow}c_{ k,\downarrow}^{\dag}\rangle=\langle c_{-k,\uparrow}d_{\uparrow}^{\dag}d_{\downarrow}c_{-k,\downarrow}^{\dag}\rangle=0$. In our numerical results, we clearly see these characters.

In Fig.~\ref{Fig:kcor1}(a), we focus on one correlation function of $P_k$, $\langle c_{-k,\uparrow}d_{\uparrow}^{\dag}d_{\downarrow}c_{k,\downarrow}^{\dag}\rangle$ with momentum $k=0.00125\pi$ in the first band below Dirac point. We do the simulation varying $\lambda$ and $U$, and we see that with large Coulomb interaction $U=5.4t$, in a wide regime, increasing the SOC slightly suppresses this correlation, this behavior agrees with that of transverse strength $J^{\parallel}$ in s-d exchange Hamiltonian $\tilde{H}$ in Eq.~(\ref{Eq:sdmodel}) (shown in Fig.~\ref{Fig:sd}(b)) in the large-$U$ limit. While with small and medium $U$, in the regime $0<\lambda<0.1t$, increasing the SOC enhances the spin-flip process.  We attribute this point to the double occupancy decreased by the SOC, consequently, the local moment of impurity is developed. In Fig.~\ref{Fig:kcor1}(b), we show the results of $\left<n_{d\uparrow}n_{d\downarrow}\right>$ as a function of $\lambda$ with different values for $U$, in particular, at $U=5.4t$, $\left<n_{d\uparrow}n_{d\downarrow}\right>$ is close to zero. We comment that in the large-$U$ limit, because the double occupancy is zero and local moment has been well developed, the backscattering with spin flip is mainly controlled by the properties for $V^{a0}_{kl\sigma}$ of Kane-Mele Hamiltonian, while with small or medium Coulomb interaction, the double occupancy dominates.

\begin{figure}
\begin{center}
\includegraphics[scale=0.38, bb=2 263 641 462]{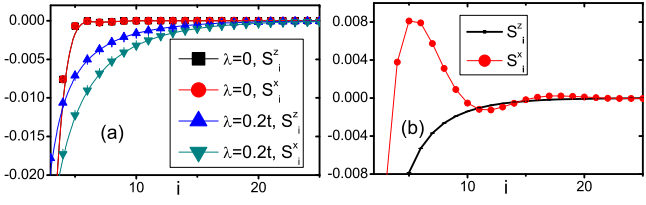}
\end{center}
\caption{(color online). Spin-spin correlation functions $S^{z}_{\mathbf{i}}$ and $S^{x}_{\mathbf{i}}$
versus the index $\mathbf{i}$ on the edge, $U=1.2t=-2\varepsilon_{d}$, and $\beta=32t^{-1}$. In (a) $\mu=0$ and in (b) $\mu=-0.2t$ and $\lambda=0.15t$.  }\label{Fig:cor}
\end{figure}

Using the same algorithm as those used for computing $P_k$, we computed the spatial distribution of Kondo cloud described by spin-spin correlation functions, $S^{z}_{\mathbf{i}}$,  $S^{x}_{\mathbf{i}}$ and $S^{y}_{\mathbf{i}}$ defined as
$S^{z}_{\mathbf{i}}=\langle(d^{\dag}_{\uparrow}d_{\uparrow}-d^{\dag}_{\downarrow}d_{\downarrow})
(c^{\dag}_{\mathbf{i}\uparrow}c_{\mathbf{i}\uparrow}-c^{\dag}_{\mathbf{i}\downarrow}c_{\mathbf{i}\downarrow})\rangle$,
$S^{x}_{\mathbf{i}}=\langle(d^{\dag}_{\uparrow}d_{\downarrow}+d^{\dag}_{\downarrow}d_{\uparrow})
(c^{\dag}_{\mathbf{i}\uparrow}c_{\mathbf{i}\downarrow}+c^{\dag}_{\mathbf{i}\downarrow}c_{\mathbf{i}\uparrow})\rangle$,
and $S^{y}_{\mathbf{i}}=(-i)^2\langle(d^{\dag}_{\uparrow}d_{\downarrow}-d^{\dag}_{\downarrow}d_{\uparrow})
(c^{\dag}_{\mathbf{i}\uparrow}c_{\mathbf{i}\downarrow}-c^{\dag}_{\mathbf{i}\downarrow}c_{\mathbf{i}\uparrow})\rangle$, and $\mathbf{i}$ is the
index of A-sublattice on the edge. In these correlation functions, we find that the spin rotational symmetry is partly broken by the SOC, i.e., $S^{z}_{\mathbf{i}}\neq S^{x}_{\mathbf{i}}$ but $S^{y}_{\mathbf{i}}= S^{x}_{\mathbf{i}}$, and this asymmetry
agrees with what was found for an impurity in the two-dimensional helical metal. \cite{Feng10} In Fig.~\ref{Fig:cor}(a), we show $S^{z(x)}_{\mathbf{i}}$ in the case of $\mu=0$, and it is clear that at $\lambda=0$, $S^{z}_{\mathbf{i}}=S^{x}_{\mathbf{i}}$ while when $\lambda\neq0$, this symmetry is broken, furthermore at $\mu=0$, both $S^{z}_{\mathbf{i}}$ and $S^{x}_{\mathbf{i}}$ has no oscillation behaviors. In Fig.~\ref{Fig:cor}(b), in the case of $\mu\neq0$ and $\lambda\neq0$, $S^{z}_{\mathbf{i}}$ still has no oscillation as \[
S^{z}_{\mathbf{i}}\propto \frac{-1}{f(R_{\mathbf{i}})},
\]
but $S^{x}_{\mathbf{i}}$ has decay as well as oscillation
\[
S^{x}_{\mathbf{i}}\propto \frac{-1}{g(R_{\mathbf{i}})}\cos2\phi R_{\mathbf{i}}.
\]
$f(R_{\mathbf{i}})$ and $g(R_{\mathbf{i}})$ are two monotonically decreasing functions of distance from impurity.

\begin{figure}[t]
\begin{center}
\includegraphics[scale=0.4, bb=39 218 647 461]{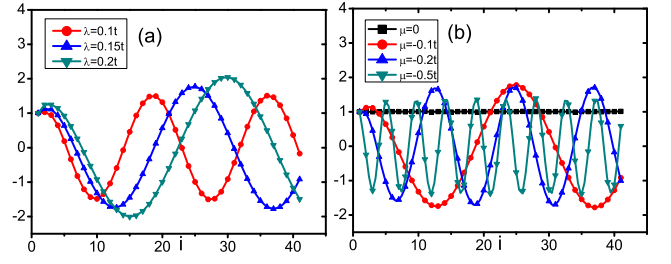}
\end{center}
\caption{(color online). (a) The ratio $S^{x}_{\mathbf{i}}/S^{z}_{\mathbf{i}}$ with $\mu=-0.1t$ . (b) The ratio $S^{x}_{\mathbf{i}}/S^{z}_{\mathbf{i}}$ with $\lambda=0.15t$ varied. In both two figures $U=1.2t=-2\varepsilon_{d}$, and $\beta=32t^{-1}$.}\label{Fig:ratio}
\end{figure}

\begin{figure}
\begin{center}
\includegraphics[scale=0.4, bb=13 231 620 461]{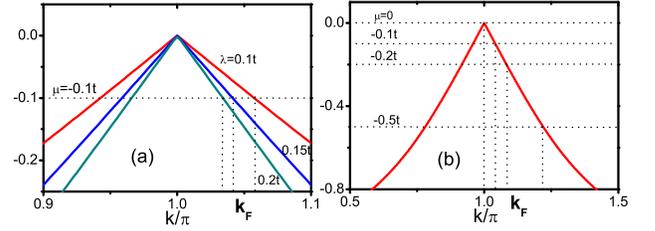}
\end{center}
\caption{(color online). Schematics of extracting Fermi vectors $k_{F}$ from the band of $H_{\textrm{K-M}}$.  (a) $\mu$ is fixed at $-0.1t$ and $\lambda=0.1t, 0.15t$ and $0.2t$. (b) $\lambda$ is fixed at $0.15t$ and $\mu=0, -0.1t, -0.2t$ and $-0.5t$. Solid lines represents the bands for Kane-Mele Hamiltonian.}\label{Fig:kf}
\end{figure}

\begin{figure}
\begin{center}
\includegraphics[scale=0.4, bb=33 252 543 472]{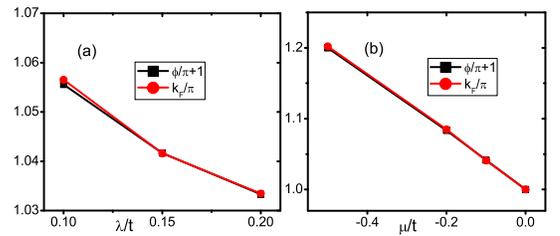}
\end{center}
\caption{(color online). Comparing $\phi$ to the Fermi vectors $k_F$. (a) is for figures~\ref{Fig:ratio}(a) and \ref{Fig:kf}(a) and (b) is for figures~\ref{Fig:ratio}(b) and \ref{Fig:kf}(b).}\label{Fig:phi_kf}
\end{figure}

\begin{figure}
\begin{center}
\includegraphics[scale=0.33, bb=58 72 534 434]{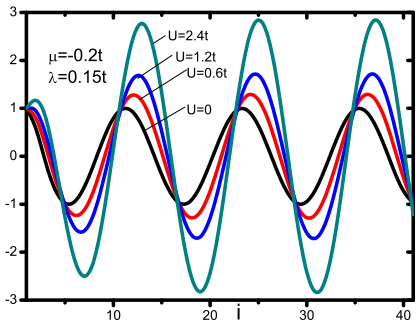}
\end{center}
\caption{(color online). The ratio $S^{x}_{\mathbf{i}}/S^{z}_{\mathbf{i}}$ in the case of $\mu=-0.2t$ and $\lambda=0.15t$ with $U$ varied. $\beta=32t^{-1}$ and $\varepsilon_{d}$ is fixed at $-U/2$. }\label{Fig:ratio_U}
\end{figure}

In order to display the asymmetry in Kondo cloud more visible, we calculate the ratio $S^{x}_{\mathbf{i}}/S^{z}_{\mathbf{i}}$ . In figures~\ref{Fig:ratio}(a) and (b), we show this ratio influenced by $\lambda$ and $\mu$, respectively. In both two cases, the difference between $S^{x}_{\mathbf{i}}$ and $S^{z}_{\mathbf{i}}$ is an oscillating factor, whose maximum and minimum is a constant, so we can conclude that $S^{z}_{\mathbf{i}}$ and $S^{x}_{\mathbf{i}}$ have the same decay part, $f(R_{\mathbf{i}}) \propto g(R_{\mathbf{i}})$. For the wave vector $\phi$ of oscillation part, Fig.~\ref{Fig:ratio} shows that $\phi$ increases both as $\lambda$ and $\mu$ decreases. We recognize that these behaviors of $\phi$ have the similarity to that of Fermi vector $k_F$, so in Fig.~\ref{Fig:kf} we extract the $k_F$ of $H_{\textrm{K-M}}$ with the same values of $\lambda$ and $\mu$ as in Fig.~\ref{Fig:ratio}. Comparing $\phi$ to the Fermi vector $k_F$ in Fig.~\ref{Fig:phi_kf}, we see that the difference between $\phi$ and $k_F$ is just a $\pi$, so $\phi$ and $k_F$ can be regarded as the same. We can conclude that
\begin{eqnarray}\label{Eq:cor}
S^{z}_{\mathbf{i}}&\propto &-\frac{-1}{f(R_{\mathbf{i}})},\nonumber\\
S^{x}_{\mathbf{i}}&\propto &-\frac{-1}{f(R_{\mathbf{i}})} \cos2k_F R_{\mathbf{i}}.
\end{eqnarray}
In Fig.~(\ref{Fig:ratio_U}), we study the effects of $U$ on ratio $S^{x}_{\mathbf{i}}/S^{z}_{\mathbf{i}}$. It is shown that the amplitude of oscillation can be enhanced but the wave vector does not change as $U$ increases.

From Fig.~\ref{Fig:cor} to Fig.~\ref{Fig:ratio_U}, we numerically study $S^{x}_{\mathbf{i}}$ and $S^{z}_{\mathbf{i}}$ in detail.
The anisotropy in Kondo cloud described by these corrlators is originated from the fact that the transverse correlation function $S^{x}_{\mathbf{i}}$ has spin-flip process and involves the contribution from the
Kramers pair $(k\sigma,-k\bar{\sigma})$ hence display $2k_F$ oscillations on Fermi level. While in vertical one $S^{z}_{\mathbf{i}}$, there is only forward scattering within the left or right
movers, which have fixed spin orientation, so it has no oscillations.
This is purely due to the helical properties on the edge, and based on this point, we also find that the charge-charge correlation $N_{\mathbf{i}}=\langle(d^{\dag}_{\uparrow}d_{\uparrow}+d^{\dag}_{\downarrow}d_{\downarrow})
(c^{\dag}_{\mathbf{i}\uparrow}c_{\mathbf{i}\uparrow}+c^{\dag}_{\mathbf{i}\downarrow}c_{\mathbf{i}\downarrow})\rangle$ is similar to
$S^{z}_{\mathbf{i}}$. Additionally, if we calculate the correlation functions between impurity and B-sublattice sites on the edge, they have the same behavior as Eq.~(\ref{Eq:cor}).  Finally, as for the decay function $f(R_{\mathbf{i}})$, here we cannot capture its exact formula in our simulation. While in Ref.~\onlinecite{Borda07}, about an Anderson impurity coupled to a one-dimensional wire, the function $f(R_{\mathbf{i}})$ has the asymptotic behavior as $R_{\mathbf{i}}^{2}$.


\section{Discussion and conclusion}
In this paper, we study an Anderson impurity on a zigzag graphene ribbon undergoing a topological phase transition driven by a Kane-Mele type SOC.
Using QMC simulations, we investigate both the local and non-local properties of the impurity. We find that with the SOC increasing from zero, the formation of local moment is supported and the impurity behavior crosses over from a spin in a band with zero width to one in a normal metal. This is because the SOC decreases the LDOS on the edge. As for non-local properties, we clearly see the helical locking and the collaboration of backscattering and spin flip in the two-particle correlators and find a broken spin rotation symmetry in spin-spin correlation functions.

Although the intrinsic SOC of carbon atoms is weak, \cite{HH06} a strong SOC can be achieved by heavy-atom doping\cite{Castro09,Abdelouahed10,Qiao10,Weeks11,Ding11,Jiang12} and tuned by a gate bias. \cite{Chen12} The magnetic impurity could be naturally realized by
dangling $\sigma$ bonds in graphene. \cite{Cazalilla12} Several physical quantities discussed can be experimentally measured: scanning tunneling microscope (STM) can measure $A(\omega)$, and recent developments in the field of spin-polarized STM \cite{Zhou10} open the possibility to detect the spin-spin correlations.

\section{Acknowledgement}
This work was supported by Academy of Finland through its Center of Excellence (2012-2017) program. The work of JEG was supported by the US Department of Energy. We acknowledge computational resources from CSC-IT Center for Science Ltd and Aalto local cluster Triton. FMH is grateful to BCCMS for hospitality during the visit.

\end{document}